\documentstyle[aps,prd,preprint,tighten]{revtex}
\begin{document}
\draft
\title{Radiative corrections to neutrino-majoron couplings
\footnote{hep-ph/9511416 \hfill HU-TFT-95-70}}
\author{Juha T. Peltoniemi
\thanks{
Juha.Peltoniemi@Helsinki.fi} }
\address{ Research Institute for Theoretical Physics, Box 9, 00014
University of Helsinki, Finland.}
\date{\today}
\maketitle
\begin{abstract}
It is claimed that radiative corrections maintain the proportionality between
the neutrino mass and the neutrino-majoron coupling and never give rise to
enhanced
decay rates in
conventional majoron models.
The coupling of a majoron to neutrinos is calculated at one loop level in
various models, including the singlet majoron model and the Zee
model with a majoron. When the respective corrections to the mass
matrix are taken into account the would-be non-diagonal terms in the
neutrino-majoron coupling are rotated away.
It is pointed out that
the coincidence between neutrino mass matrices and neutrino-majoron couplings
is not accidental, but is a general consequence of  N\"{o}ther's theorem.
N\"{o}ther's theorem also implies that the majoron coupling to charged
fermions is always diagonal in the fermion mass basis, and it vanishes
completely in the singlet majoron models.

\end{abstract}

\section{Introduction}

Most of the models explaining naturally a small neutrino mass break
the global B-L-symmetry that is present in the Standard Model.
Here is considered the case that the symmetry is broken spontaneously,
by a non-zero vacuum expectation value of a new scalar field.
As a consequence of this, there will arise a massless Goldstone boson,
majoron ($J$).

In models with a massless majoron a massive neutrino may decay to a lighter
neutrino  state and a majoron. The life time of a neutrino is given by
\begin{equation}
\tau_\nu = \frac{32 \pi}{|\xi|^2 m_\nu},
\end{equation}
where $\xi$ is the non-diagonal majoron coupling, and $m_\nu$ is the mass of
the decaying neutrino, and the daughter neutrino is assumed to be much lighter.

If the neutrinos are expected to form the dark
matter,
or part of it, they should be sufficiently long lived, i.e. should have a life
time at least as long as the age of the universe ($t_0 \sim 10^{17}$ s).
On the other hand, a sufficiently rapid decay allows the neutrinos to have
masses above 100 eV, forbidden for stable neutrinos.
To be consistent with the present status of the universe, one requires
\cite{Dicusetal78,Pal83}
\begin{equation}\tau_\nu < t_0 h^4
\left(\frac{100 {\hspace{0.3ex}\rm eV}}{m_\nu}\right)^2,
\end{equation}
where
$h$ is the normalized Hubble constant.
The models of large scale structure formation impose limits that are several
magnitudes more
stringent, though more model dependent
\cite{SteigmanTurner85}.

Typically the neutrino-majoron coupling is
equal to the neutrino mass divided by
the
vacuum
expectation value of the scalar field. Although this coupling itself may not be
small,
the decay
amplitudes may be strongly suppressed. This is due to the fact that the
majoron coupling matrix will be be diagonalised
simultaneously with the mass
matrix (just as the standard model higgs coupling is diagonal in the mass
basis). For example, in the triplet majoron model  the
tree level majoron coupling is exactly diagonal in the mass basis.

In models where some components of the mass matrix are generated by a
different mechanism  than others,
due to the different quantum numbers of the neutrinos involved, the
coincidence of the mass and majoron coupling is broken. The simplest example
is the see-saw model \cite{GellmannRamondSlansky79,Yanagida79} with a singlet
majoron \cite{ChikashigeMohapatraPeccei81}, where only the Majorana mass of
the right-handed sector breaks the lepton symmetry, and the Dirac mass
components obey it. Thus non-diagonal majoron couplings arise at tree level,
though these may be very small due to a characteristic hierarchy of these
models \cite{SchechterValle82}.

It is not so evident that this structure is maintained beyond the tree level.
Indeed, it has been claimed \cite{GeorgiGlashowNussinov81,Pal90} that
radiative corrections would induce non-diagonal components to the
neutrino-majoron couplings.
According to  Ref.~\cite{GeorgiGlashowNussinov81},  the graph
\ref{kuvaTrip} d
should allegedly induce a sufficiently large non-diagonal coupling
to make light neutrinos
essentially unstable in the triplet majoron model.

Here I will show that the appearance of non-diagonal elements is not true.
Since we are actually interested in the form of the majoron
coupling in the true mass basis,
it is not enough to consider the corrections to the majoron
coupling, but one should also take into account
the respective corrections into the mass matrix.
It will turn out that the mass eigenbasis is also rotated
in such a way that the majoron coupling returns to
the diagonal form.
This is shown generally, in a model independent way, using
the conservation of currents. To be sure, it is also
proven explicitly in several models, by calculating the loops.
Previously it has been shown that in a simplified
singlet majoron model the radiative corrections do
not enhance the decay \cite{ClineKainulainenPaban93}.

In this paper I will calculate
at one loop level the neutrino-majoron
coupling in the singlet majoron model
explicitly, more generally than in
\cite{ClineKainulainenPaban93}.
It is found that radiative corrections to the mass matrix
coincide with the corrections to the majoron couplings,
so that the neutrino decay is described sufficiently
well by the tree level result.
It is shown also
that at one loop level the neutrino-majoron
coupling is exactly diagonal (in the mass eigenbasis)
in the majoronic version of the Zee model  \cite{Zee80,Zee85}.
It is also argued that the results apply to other similar models,
like the triplet, doublet \cite{BertoliniSantamaria88,KikuchiMa94},
singlet-triplet, doublet-triplet
or singlet-doublet-triplet majoron models, as well as other
radiative models.
It is evident that this is true also for more complicated models where the
majoron is incorporated in some Susy or GUT scenario whose low
energy characteristics resemble the models considered here.

It has also been widely believed that radiative corrections would
induce a coupling of a majoron to charged leptons, even when the
tree level coupling would be zero. Because of these couplings
the heavy leptons might then decay as $\ell_2^- \to \ell_1^- J$.
However, the current argument implies that the radiative corrections do not
induce any
non-diagonal couplings, and in models where the majoron is a singlet
the majoron does not couple at all to any charged fermion.

\section{General approaches}\label{Noether}

In Ref.\ \cite{ClineKainulainenPaban93} it was claimed that the diagonality
of the majoron coupling is a direct consequence of the renormalisation
procedure. According to this principle,
any results obtained at the
symmetric phase, i.e. for a zero vacuum expectation value, should be
analytically continuable to the phase where this symmetry is hidden due to
the arising non-zero vacuum expectation value. Hence the
coincidence of mass and scalar couplings should be manifest.
However, this kind of treatment is
valid only for renormalisable dimension 4 terms.
Higher dimension terms can arise already at one loop level.

Another approach based on N\"{o}ther's theorem
was presented in
Ref.~\cite{BertoliniSantamaria88}, for the doublet majoron model.
Below this is generalized to an arbitrary model.

Majoron models always contain two U(1) symmetries, one associated with the
hypercharge Y and the other with the lepton number L. For each symmetry
one can define a current
\begin{eqnarray}\label{JL}
J_\mu^L &=& \sum_h L_h \Phi_h^\dagger \partial_\mu \Phi_h
+ \sum_f L_f \bar{\Psi}_f \gamma_\mu \Psi_f,
\\ \label{JY}
J_\mu^Y &=& \sum_h Y_h \Phi_h^\dagger \partial_\mu \Phi_h
+ \sum_f Y_f \bar{\Psi}_f \gamma_\mu \Psi_f,
\end{eqnarray}
where $h$ runs over scalar states and $f$ over fermion states, and
$L_i$ and $Y_i$ label the lepton number and the hypercharge of particle $i$,
respectively.
Whenever the symmetry is exact these currents are conserved, i.e.
\begin{equation}\label{Noet}
\partial^\mu J_\mu^{L,Y} = 0.
\end{equation}
Furthermore, the currents are conserved also after the spontaneous
breaking of the symmetry, as long as the underlying theory itself
is renormalisable and anomaly free. However, anomalies break
N\"{o}ther's theorem by instanton effects, and for exactly this
reason we should expand the lepton number symmetry to
include baryons ($B-L$).

In the phase of spontaneously broken symmetries we can shift the
neutral components by their vacuum expectation values, so that
the equations (\ref{JL}) and (\ref{JY}) can be written as
\begin{eqnarray}\label{JLb}
J_\mu^L &=&
\sum_h L_h \langle \Phi_h \rangle \partial_\mu {\rm Im}\{ \Phi_h\}
+ j_\mu^L
\\ \label{JYb}
J_\mu^Y &=&
\sum_h Y_h \langle \Phi_h \rangle \partial_\mu {\rm Im}\{\Phi_h \}
+ j_\mu^Y,
\end{eqnarray}
where the $j_\mu$ include the other terms of Eqs.~(\ref{JL}) and (\ref{JY}).
The first order terms in (\ref{JYb}) consist of the unphysical Goldstone mode
associated
with the $Z$ boson, by definition, while those in (\ref{JLb}) include both
the unphysical higgs and the true Goldstone boson, majoron.
The majoron is defined as
\begin{equation}
J= C \sqrt{2}
\sum_h \left( \sum_i Y_i^2 v_i^2 L_h - \sum_i L_i Y_i v_i^2 Y_h
\right) v_h {\rm Im}\{ \Phi_h \},
\end{equation}
where
\begin{equation}
C^{-1} =
{\sqrt{\sum_h \left( \sum_i Y_i^2 v_i^2 L_h - \sum_i L_i Y_i v_i^2 Y_h
\right)^2 v_h^2}}.
\end{equation}
Arranging the terms, using the conservation of currents (\ref{Noet}), and
eliminating
the would-be Goldstone mode of $Z$, one obtains
\begin{eqnarray}
\partial^\mu \partial_\mu J +
2 C \partial^\mu
 ( \sum_i Y_i^2 v_i^2 j_\mu^L - \sum_i L_i Y_i v_i^2 j_\mu^Y  )
 = 0.
\end{eqnarray}
Now one can see that the
neutrino-majoron coupling should be of the form
\begin{eqnarray}\label{nemajex}
{\cal L}_J &=&
C\left[ \left(\sum_h Y_h^2 v_h^2\right)
\left( \sum_f U_{af}^\dagger L_f U_{fb}\right)
\right. \nonumber \\&& \mbox{} \left.
- \left(\sum_h L_h Y_h v_h^2\right)
\left( \sum_f U_{af}^\dagger Y_f U_{fb}\right) \right]
J \bar{\nu}_a (\stackrel{\leftarrow}{\partial}_\mu -
\stackrel{\rightarrow}{\partial}_\mu) \gamma^\mu
\nu_b .
\end{eqnarray}
Using the Dirac equation (for Majorana neutrinos) one can extract the coupling
constant
\begin{eqnarray}\label{nemaj}
\xi_{ab} &=&
 C\left(\sum_h Y_h^2 v_h^2\right)
 \left( \sum_f U_{af}^\dagger L_f U_{fb}\right)(m_a+m_b)
 \nonumber \\&& \mbox{}
- C\left(\sum_h L_h Y_h v_h^2\right)
\left( \sum_f U_{af}^\dagger Y_f U_{fb}\right)(m_a+m_b),
\end{eqnarray}
where $m_{a,b}$ are Majorana masses of neutrinos a,b.
(For Dirac neutrinos the result is different.)
One sees immediately that a necessary and sufficient condition
for the majoron coupling being diagonal in the neutrino mass basis
is that all neutrinos are identical in their hypercharge and lepton numbers

To be sure that the above reasoning is correct, below it will be proven
explicitly that one loop corrections give results in full agreement with
the above results, in the singlet majoron model with the see-saw mechanism
and the Zee model.

\section{Singlet majoron model}

\subsection{Tree level }

The singlet majoron model
\cite{ChikashigeMohapatraPeccei80} is the most evident
extension to the standard model
containing a majoron.
This model involves, beside the Standard Model particles, three right handed
neutrinos, with a canonical lepton number 1, and a neutral scalar
$\sigma$ with
lepton number~2.
The Yukawa interactions of neutrinos are given by
\begin{equation}
 \Phi^0 \bar{\nu}_L f \nu_R  +  \sigma \bar{\nu}_R h \nu_R,
\end{equation}
where $\Phi$ is the standard model higgs doublet,
and $\nu_L$ and $\nu_R$ are 3
dimensional vectors representing the three neutrino flavors,
and the coupling constants $f$ and $h$ are
respectively $3\times 3$ matrices.

After the spontaneous symmetry breaking by a non-zero vacuum expectation
value of $\sigma$,
\begin{equation}\label{vev}
\langle \sigma \rangle \equiv
\frac{w}{\sqrt{2}},
\end{equation}
there will be a massless majoron, $J$, as the imaginary part of the field,
\begin{equation}
\sigma = \frac{1}{\sqrt{2}} \left( w + \rho + iJ \right).
\end{equation}
The standard model higgs can also be decomposed as
\begin{equation}
\Phi^0 = \frac{1}{\sqrt{2}} \left( v + H^0 + i\phi^0 \right),
\end{equation}
with $\langle \Phi^0 \rangle = v/\sqrt{2}$.
A cross coupling of the scalars will induce a mixing between the
real parts. Define the mass eigenstates as
\begin{eqnarray}
H_1 &=& \cos\theta  H^0 - \sin\theta  \rho, \\
H_2 &=& \sin\theta  H^0 + \cos\theta  \rho,
\end{eqnarray}
with masses $M_1$ and $M_2$.
The Goldstone modes do not mix, as $\sigma$ carries no
hypercharge, and $\Phi$ has no lepton number.

The tree level neutrino mass matrix reads now
\begin{equation}
{\cal M}_\circ = \frac{1}{\sqrt{2}}\left[
\begin{array}{cc}
0            & fv \\
f^\dagger v & hw
\end{array}
\right]
\equiv
\left[
\begin{array}{cc}
0 & M_D \\
M_D^\dagger & M_M
\end{array}
\right]
\end{equation}
where $M_D$ and $M_M$ are $3\times 3$ matrices. This can be block diagonalised
by the matrix \cite{MohapatraPal88,Kanaya80}
\begin{equation}
U = \left[ \begin{array}{cc}
1- \frac{1}{2} \epsilon\epsilon^\dagger
   &  \epsilon - \frac{3}{2}\epsilon\epsilon^\dagger\epsilon \\
\epsilon^\dagger + \frac{3}{2}\epsilon^\dagger\epsilon\epsilon^\dagger
   &   1 - \frac{1}{2} \epsilon^\dagger \epsilon
\end{array}\right]
\end{equation}
to the order of $\epsilon^4$, where
\begin{equation}
\epsilon = M_D M_M^{-1}.
\end{equation}
After the rotation we obtain the mass matrix
\begin{equation}
{\cal M}_\circ = \left[
\begin{array}{cc}
\widehat{m} & 0 \\
0 & \breve{m}
\end{array}
\right] +
\left[
\begin{array}{cc}
O(\epsilon^6) & O(\epsilon^5) \\
O(\epsilon^5) & O(\epsilon^4)
\end{array}
\right]      ,
\end{equation}
where
\begin{eqnarray}\label{treeMl}
\widehat{m} &=&-\epsilon M_M \epsilon^\dagger
+ \frac{1}{2} (
\epsilon M_M\epsilon^\dagger\epsilon\epsilon^\dagger
+\epsilon \epsilon^\dagger\epsilon M_M\epsilon^\dagger),
\\
\breve{m} &=& M_M  + \frac{1}{2}( M_M \epsilon^\dagger \epsilon
+ \epsilon^\dagger \epsilon  M_M ).
\end{eqnarray}
The remaining submatrix of the light neutrinos can be diagonalised by a further
rotation in that subpart by a ($3 \times 3$) unitary matrix.

Define now the majoron coupling as
\begin{equation}
-i \xi_{ij}   J \bar{\nu}^c_i \gamma_5\nu_j.
\end{equation}
One sees that at the tree level we have $\xi = h/\sqrt{2} = M_M/w$, and
in the weak basis only the $3\times 3$ submatrix of the right-handed states in
$\xi$ is non-zero. After the block diagonalisation one has
\begin{eqnarray}
\xi &=& \frac{1}{w}\left[
\begin{array}{cc}
\widehat{\xi}& -\epsilon M_M  \\
- M_M \epsilon^\dagger & \breve{\xi}
\end{array}
\right]
+ \left[\begin{array}{cc}
O(\epsilon^6) & O(\epsilon^3) \\
O(\epsilon^3) & O(\epsilon^4)
\end{array}\right]
\end{eqnarray}
where
\begin{eqnarray}
\widehat{\xi} &=& \epsilon M_M\epsilon^\dagger -
\frac{3}{2}(\epsilon M_M \epsilon^\dagger \epsilon\epsilon^\dagger
+ \epsilon\epsilon^\dagger \epsilon M_M \epsilon^\dagger )
\\
\breve{\xi} &=& M_M  - \frac{1}{2}( M_M \epsilon^\dagger \epsilon
+ \epsilon^\dagger \epsilon  M_M )
\end{eqnarray}
The elements for the coupling between the heavy and the light
mass eigenstates are substantial, but the couplings of both the heavy and the
light mass
eigenstates are in themselves separately diagonal in the mass basis, at
$O(\epsilon^3)$.
As a result of this, the decay of a light massive neutrino to a lighter
neutrino
and a majoron is suppressed by $\epsilon^4$ at the tree level.
The leading order
for neutrino decay is now $\epsilon^4$,
the respective couplings can be written
as \cite{MohapatraPal88}
\begin{equation}\label{mpeq}
\widehat{\xi'} = - \frac{1}{w} \left( \widehat{m} \epsilon \epsilon^\dagger +
\epsilon \epsilon^\dagger \widehat{m}
\right).
\end{equation}
For all plausible values of the parameters, this is too small to
induce a rapid neutrino decay.

\subsection{Radiative corrections}

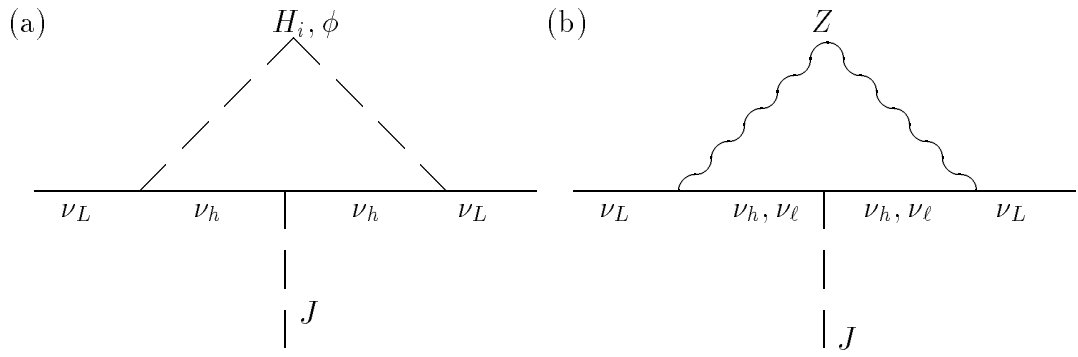
\begin{figure}
\caption{The diagrams generating the corrections to the majoron coupling with
the
left-handed neutrino
states in the singlet majoron model.
}
\label{kuvaLLJ}
\end{figure}

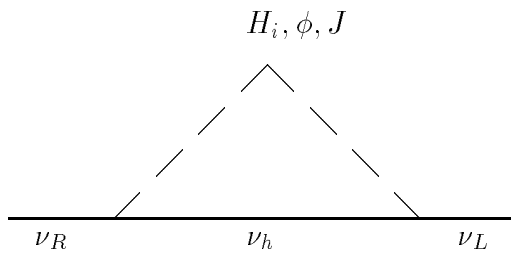
\begin{figure}
\caption{The diagrams contributing to the Dirac mass term in the singlet
majoron
model
}
\label{kuvaLRm}
\end{figure}

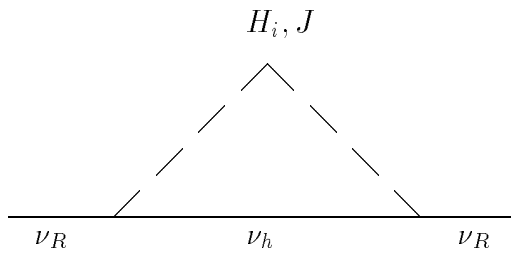
\begin{figure}
\caption{The diagrams giving the correction to the mass term for the
right-handed states in the singlet majoron model.
}
\label{kuvaRRm}
\end{figure}

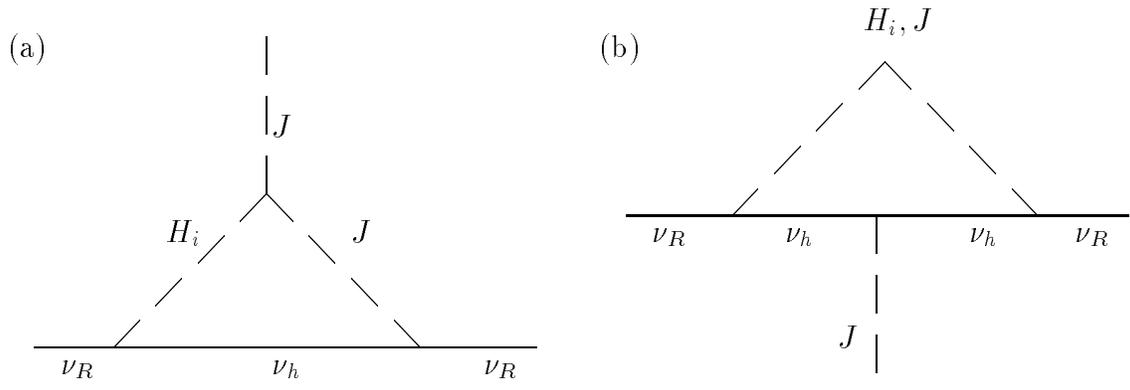
\begin{figure}
\caption{The diagrams generating the lowest order corrections to the majoron
coupling between the
right-handed states in the singlet majoron model.
}
\label{kuvaRRJ}
\end{figure}

Calculate now exactly the most important radiative corrections to the mass and
majoron coupling.
The graphs contributing to the mass are presented in figs.\
\ref{kuvaLRm} and \ref{kuvaRRm}
while the graphs with an external majoron are depicted in figs.\
\ref{kuvaLLJ}, \ref{kuvaLRJ} and \ref{kuvaRRJ}.
To obtain maximum clarity,
we do our calculations in the weak (chirality) basis
of the
external states, labeling the left-handed states
with $i,j$ and the right-handed
states by $a,b$.
The readers interested in a more elaborate
treatment in the tree level mass basis with further rerotations into
the corrected mass basis should consult the ref.\
\cite{ClineKainulainenPaban93}.
The internal particles are always considered to be mass eigenstates.
Moreover, at the desired accuracy we can assume that the
heavy mass eigenstates coincide with the right-handed states.
Hence the $\sigma$
coupling can be taken to be diagonal.
All the loop
calculations are
performed in the Feynman-t'Hooft gauge.

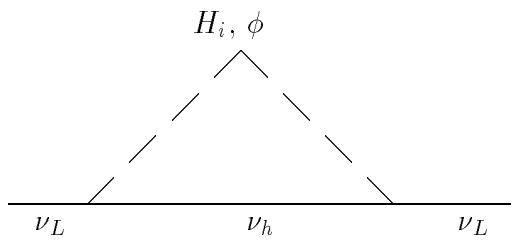
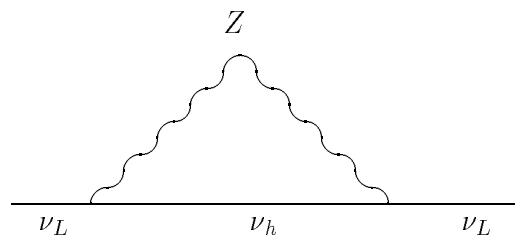
\begin{figure}
\caption{The diagrams generating the mass correction to the left-handed
states in the singlet majoron model.
}
\label{kuvaLLm}
\end{figure}

Let us start with the corrections to the {\em left-handed} states.
The
graphs contributing to the mass matrix are
presented in  figure  \ref{kuvaLLm}.
The two graphs containing scalars diverge separately, but
their sum is finite, giving
\begin{eqnarray}\label{mac}
\delta m_{ij}^{(H_i,\phi)}
 &=& \frac{1}{32\pi^2} \sum_a f_{ia} \check{m}_a f_{ja}
 \nonumber \\ &&
\left[ (M_1^2-M_Z^2) I(\check{m}_a,M_1,M_Z)
\frac{ }{ }\right.\nonumber \\&&\mbox{}\left.
- \sin^2\theta (M_2^2-M_1^2) I(\check{m}_a, M_1, M_2)
\frac{ }{ } \right],
\end{eqnarray}
where
\begin{eqnarray} \label{I}
\lefteqn{I(m_1, m_2, m_3 ) = } \\ & &
\frac{m_1^2 m_2^2 \ln\frac{m_1^2}{m_2^2}
+m_2^2 m_3^2 \ln\frac{m_2^2}{m_3^2}
+m_3^2 m_1^2 \ln\frac{m_3^2}{m_1^2}
}{(m_2^2-m_1^2)(m_3^2-m_2^2)(m_1^2-m_3^2)},
\nonumber
\end{eqnarray}
and $\theta$ is the scalar mixing angle.
Respectively, the scalar graphs in figure \
ref{kuvaLLJ} yield the majoron
coupling
\begin{eqnarray} \label{jc}
\delta \xi_{ij}^{(H_i,\phi)}     &=&
 \frac{1}{32\sqrt{2} \pi^2} \sum_a f_{ia} h_a f_{ja}
 \nonumber \\ &&
\left[ (M_1^2-M_Z^2) I(\check{m}_a,M_1,M_Z)
\right.\nonumber \\ && \mbox{ }\left.
- \sin^2\theta (M_2^2-M_1^2) I(\check{m}_a, M_1, M_2)
\frac{ }{ } \right]
\end{eqnarray}
The graph with $Z$ exchange (\ref{kuvaLLm} b) requires a more careful
treatment.
It is now compulsory to take all the neutrino states into account as the
internal lines: ignoring the light states would leave a diverging integral.
{}From the definition of
$\epsilon$ one sees that these two sets of graphs are indeed of the same
order in $\epsilon$, and can be summed. In the lowest order of $\epsilon$ one
obtains
\begin{equation}\label{macc}
\delta m^{(Z)}_{ij} = \frac{g^2}{256\pi^2\cos^2\theta_W}
\sum_a \epsilon_{ia} \check{m}_a^3 I(\check{m}_a,M_Z,0) \epsilon^\dagger_{aj}.
\end{equation}
The $Z$ contribution to the majoron coupling (Fig.\ \ref{kuvaLLJ} b) is
respectively
\begin{eqnarray}
\delta \xi^{(Z)}_{ij} &=& \frac{g^2}{256\sqrt{2}\pi^2\cos^2\theta_W}
\nonumber \\ & &
\sum_a \epsilon_{ia} \check{m}_a^2 h_a I(\check{m}_a,M_Z,0)
\epsilon^\dagger_{aj}.
\end{eqnarray}
where the equation (\ref{treeMl}) was used.

Consider then the {\em left-right} components.
The Dirac mass term is corrected by the graph \ref{kuvaLRm}.
We obtain
\begin{eqnarray}
\delta m_{ia} &=& \sum_{bj}
\frac{1}{32\pi^2} f_{ib} \check{m}_b \epsilon_{bj}^\dagger f_{ja}
\nonumber \\&&
\left[ \cos^2\theta I(\check{m}_b,M_1,M_Z)
+ \sin^2\theta I(\check{m}_b,M_2,M_Z)
\right].
\end{eqnarray}
At tree level there was no coupling of a majoron between a left- and a
right-handed neutrino. At one loop level, the graphs presented in
figures \ref{kuvaLRJ} would contribute to such a coupling.
However, it turns out that these graphs cancel so that there is no
respective majoron coupling at all. This is shown in the
Appendix.

Continue then with the components involving only the {\em right-handed}
neutrino states. The
one-loop correction to
the mass term, from figure \ref{kuvaRRm}, can be expressed as
\begin{eqnarray}
\delta m_{aa}&=& \frac{1}{32\pi^2} h_a^2 \check{m}_a
\left[ \cos^2\theta M_2^2 I(\check{m}_a,M_2,0) \right.
\nonumber \\ & & \mbox{ } \left.
+ \sin^2\theta M_1^2 I(\check{m}_a,M_1,0)\right].
\end{eqnarray}
Other contributions appear at higher order of $\epsilon$, being thus
ignorable.

There are now four graphs contributing to the majoron coupling.
The graph (\ref{kuvaRRJ}a),
with the majoron coupling to the scalar vertex gives
\begin{eqnarray}
\delta \xi_a^{(\rho J)} &=& \frac{1}{32\pi^2}  h_a^2 \frac{\check{m}_a}{w}
\left[ M_2^2\cos^2\theta I(\check{m}_a,M_2,0)
\right.\nonumber \\& & \mbox{ }\left.
+ M_1^2\sin^2\theta I(\check{m}_a,M_1,0)\right].
\end{eqnarray}
There is another graph with $ \rho$ and $J$ lines interchanged which gives
exactly the
same contribution.
The graphs (\ref{kuvaRRJ}b) with the majoron coupling to the fermion line
diverge
separately, but together they sum up to the finite result
\begin{eqnarray}
\delta \xi_{a}^{(\nu)} &=& - \frac{1}{32\sqrt{2}\pi^2} h_a^3
\left[ \cos^2 \theta M_2^2
I(\check{m}_a,M_2,0)
\right.\nonumber \\& & \mbox{  } \left.
+ \sin^2 \theta M_1^2 I(\check{m}_a,M_1,0) \right].
\end{eqnarray}
The rest of the contributing graphs can be ignored since
they are of higher order in $\epsilon$.

Summing up all contributions, one sees that the
one loop contributions to the majoron couplings
are given by
\begin{equation}
\delta \xi = \frac{1}{w}
\left[ \begin{array}{cc}
\delta m_L & 0 \\
0              & \delta m_R
\end{array}\right].
\end{equation}
{}From this it follows that after diagonalising the mass matrix,
by a redefined rotation matrix, the resulting majoron coupling
is still diagonal in the lowest order, and the non-diagonal
majoron couplings are obtained from the same equation
(\ref{mpeq}) as in the tree level case (with only the
quantities $m$ and $\epsilon$ involving the tiny loop
corrections).

\section{Radiative models}
\subsection{Majoronic Zee model}

The model of Zee \cite{Zee80,Zee85} is the simplest mechanism to generate
the neutrino
masses radiatively. This model is often incorporated with a global symmetry,
which can be a flavor dependent Zeldovich-Konopinski-Mahmoud symmetry
\cite{Zeldovich52,KonopinskiMahmoud53} or a modification of it
\cite{Petcov82,Valle85}. Here I restrict on the variant where this symmetry is
the canonical lepton number symmetry,
whose spontaneous breakdown then produces
the majoron.

There are several ways to insert the majoron to the Zee model.
The simplest is to assume the majoron to be composed of the
doublets present in the Zee model \cite{BertoliniSantamaria88}.
This variant is, however, incompatible with the LEP results, so
one has to add more higgses. In the following is considered the
case that the majoron emerges from a singlet field. Another
possibility would be to add a third doublet, in the same spirit
as in the model of Ref.\ \cite{KikuchiMa94}. The final results
are equal for all cases, up to a trivial scalar mixing factor in the
majoron couplings.

The considered model
has the minimal Standard Model fermion sector, including no right-handed
neutrinos, with canonical lepton numbers. The higgs sector consists of the
standard
higgs doublet $\Phi$,
 one singly charged singlet scalar $\eta^-$, one additional doublet
$\varphi$, and a neutral scalar $\sigma$. Both of the singlets have lepton
number 2 while the doublets do not carry any lepton number.

Only one of the doublets couples to leptons, and the Yukawa couplings
to neutrinos are given by
\begin{equation}
h_i \Phi^+ \bar{\nu}_{Li} l_{Ri}^- + f_{i j} \eta^- {\bar
l}^-_{Li}
\nu_{Lj},
\end{equation}
where $f_{ij}$ is antisymmetric.
The $\sigma$ field couples only to the scalars,  with a quartic coupling
\begin{equation}
\lambda \Phi^T \varphi \eta^- \sigma
\end{equation}
where $\lambda$ is a dimensionless coupling constant.

After the spontaneous symmetry breakings we have 3 charged scalars, one of
which is eaten up by $W$. It is
sufficient to describe their mixing by two mixing angles: $\alpha$ is the
mixing between physical and unphysical components,
\begin{equation}
\tan \alpha = \frac{\langle \Phi \rangle }{ \langle \varphi \rangle},
\end{equation}
and $\beta$
is the mixing between the physical eigenstates,
\begin{equation}
\tan 2\beta = \frac{4 \sqrt{2} \lambda \langle \sigma \rangle }{g }
\frac{M_W}{M_\phi^2-M_\eta^2 },
\end{equation}
where $M_\phi$ and $M_\eta$ are the diagonal mass terms in the weak basis.
One sees easily that the coupling of the majoron to the physical charged bosons
is
purely non-diagonal in the mass basis, the respective coupling constant being
$i\lambda \sqrt{\langle \Phi \rangle^2 + \langle \varphi \rangle^2 }/
\sqrt{2}$.

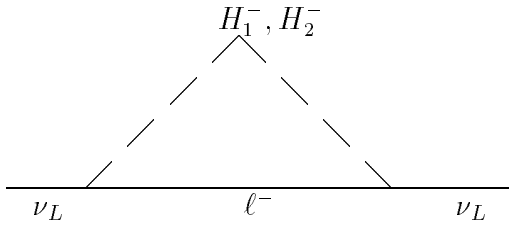
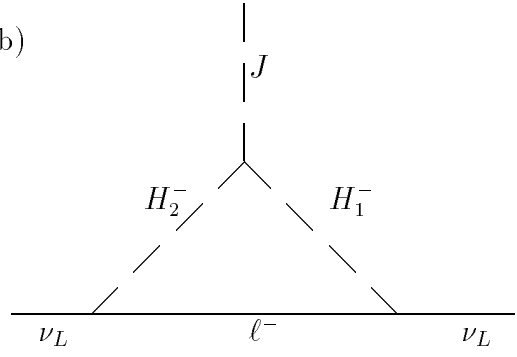
\begin{figure}
\caption{The diagrams generating the mass and the majoron coupling in the Zee
model with singlet majoron.
}
\label{kuvaZeem}
\end{figure}

The mass is given by the graph (a)
of the figure \ref{kuvaZeem} \cite{Petcov82}
\begin{eqnarray}\label{zee-mass}
m_{ik} &=& \frac{f_{ik} g \sin 2\beta \cot \alpha}{32 \pi^2 M_W}
(M_2^2-M_1^2)
\nonumber \\&&
\left[  I(m_i, M_1, M_2) -  I(m_k, M_1, M_2) \right],
\end{eqnarray}
where the function $I$ is the same as before
(\ref{I}), and $M_1$ and $M_2$ are
the mass eigenstates.
In the limit $m_k \ll M_j$ the mass reads
\begin{equation}
m_{ik} = \frac{f_{ik} g \sin 2\beta \cot \alpha}{32 \pi^2 M_W}
(m_i^2-m_k^2) \ln\frac{M_1^2}{M_2^2}.
\end{equation}
This limit is sufficient for most practical applications.

The majoron coupling (Figure \ref{kuvaZeem} b) is respectively given by
\begin{eqnarray}\label{zee-majo}
\xi_{ik} &=&
\frac{f_{ik} g \lambda \cot \alpha
 \sqrt{\langle \Phi \rangle^2 + \langle \varphi \rangle^2 }
}{16 \sqrt{2}
\pi^2 M_W}
\nonumber \\&&
\left[  I(m_i, M_1, M_2) - I(m_k, M_1, M_2) \right].
\end{eqnarray}
Using the definition of the mixing angle and taking the heavy scalar limit
one obtains 
\begin{equation}
\xi_{ik} = \frac{f_{ik} g \sin 2\beta \cot \alpha}{32 \pi^2 w M_W}
(m_i^2-m_k^2) \ln\frac{M_1^2}{M_2^2}.
\end{equation}
Hence one has also in
this model exactly
\begin{equation}
m_{ij} = \xi_{ij} w
\end{equation}
at one loop level.

\subsection{Other radiative models}

It is obvious now that
these results can be generalized to other radiative models with similar
kind of loops.
One example of them is the model considered in \cite{Peltoniemi93}.
In it it was proposed that singlet neutrinos could be
light, with radiatively generated masses.
It can be shown that in that model
the radiatively induced neutrino-majoron
coupling is diagonal, and the neutrinos are essentially stable,
justifying the motivation for the model of having the singlet
neutrinos as dark matter.

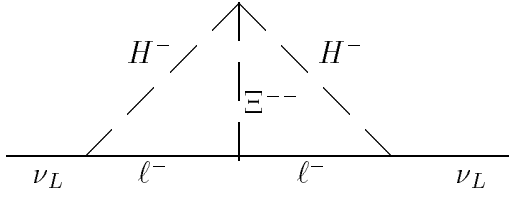
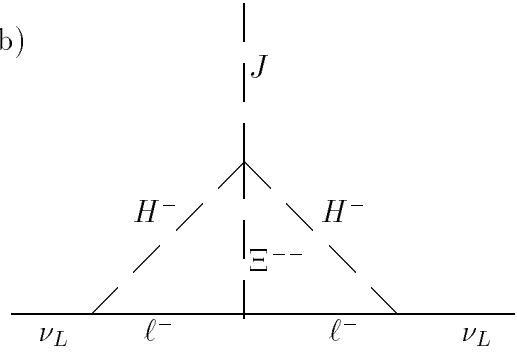
\begin{figure}
\caption{The diagrams generating the mass and the majoron coupling in the Babu
model with singlet majoron.
}
\label{kuvaBabu}
\end{figure}

A more trivial case is
the model of Babu \cite{Babu88} with a singlet majoron.
In that model the mass is
generated at two loop level, and
the majoron couples only to the charged scalar
bosons. Hence it is free from any serious constraints from lepton
phenomenology.
The mass in now given by the graph \ref{kuvaBabu}a, while the
majoron coupling results from \ref{kuvaBabu}b. These graphs
are trivially proportional to each others.

There is a plethora of other models with radiative generation
of masses. Many of them use non-canonical family dependent
lepton number
assignment or light right-handed neutrinos. In such scenarios the
neutrino decay can take place via the lowest order graphs,
and it may not be suppressed essentially.

\section{Triplet majoron model}

The triplet
majoron model \cite{GelminiRoncadelli81}
is the simplest alternative to produce
neutrino masses without right-handed neutrinos.
It was formerly very popular,
but the LEP results
were rather detrimental to this model, excluding it in its simplest forms.
The triplet scenario itself is not dead, however, it survives as a part
of more complicated models, e.g.\ the singlet-triplet model.
Hence it is well motivated to study the triplet model as
a starting point to more elaborate models.

In the triplet majoron models one adds to the minimal
standard model a triplet field $\Delta $, with hypercharge
2 and lepton number 2. It couples to lepton doublets ($\Psi_L$) via the
coupling
\begin{equation}
\zeta_{ij} \Psi_{Li}^T \Delta \Psi_{Lj}.
\end{equation}
The lepton number is now broken spontaneously by the vacuum
expectation value of $\Delta$, denoted by
$\langle \Delta \rangle \equiv u /\sqrt{2}$.
Even though there are no cross couplings between the scalars,
the gauge interactions mix the Goldstone modes, and the majoron is a linear
combination of the
imaginary parts
of both neutral higgses,
\begin{equation}
J = \sqrt{2}\; {\rm Im}\left\{\frac{2u \Phi^\circ - v \Delta^\circ}
{\sqrt{v^2+4u^2}}\right\}.
\end{equation}
The unphysical higgs field is then orthogonal to the majoron.

The neutrino-majoron coupling is now at tree level
\begin{equation}\label{xim}
\xi_0 = \frac{m_0}{u}\frac{v}{\sqrt{v^2+4u^2}}.
\end{equation}
The majoron has also a coupling to the charged fermions,
via the doublet component.

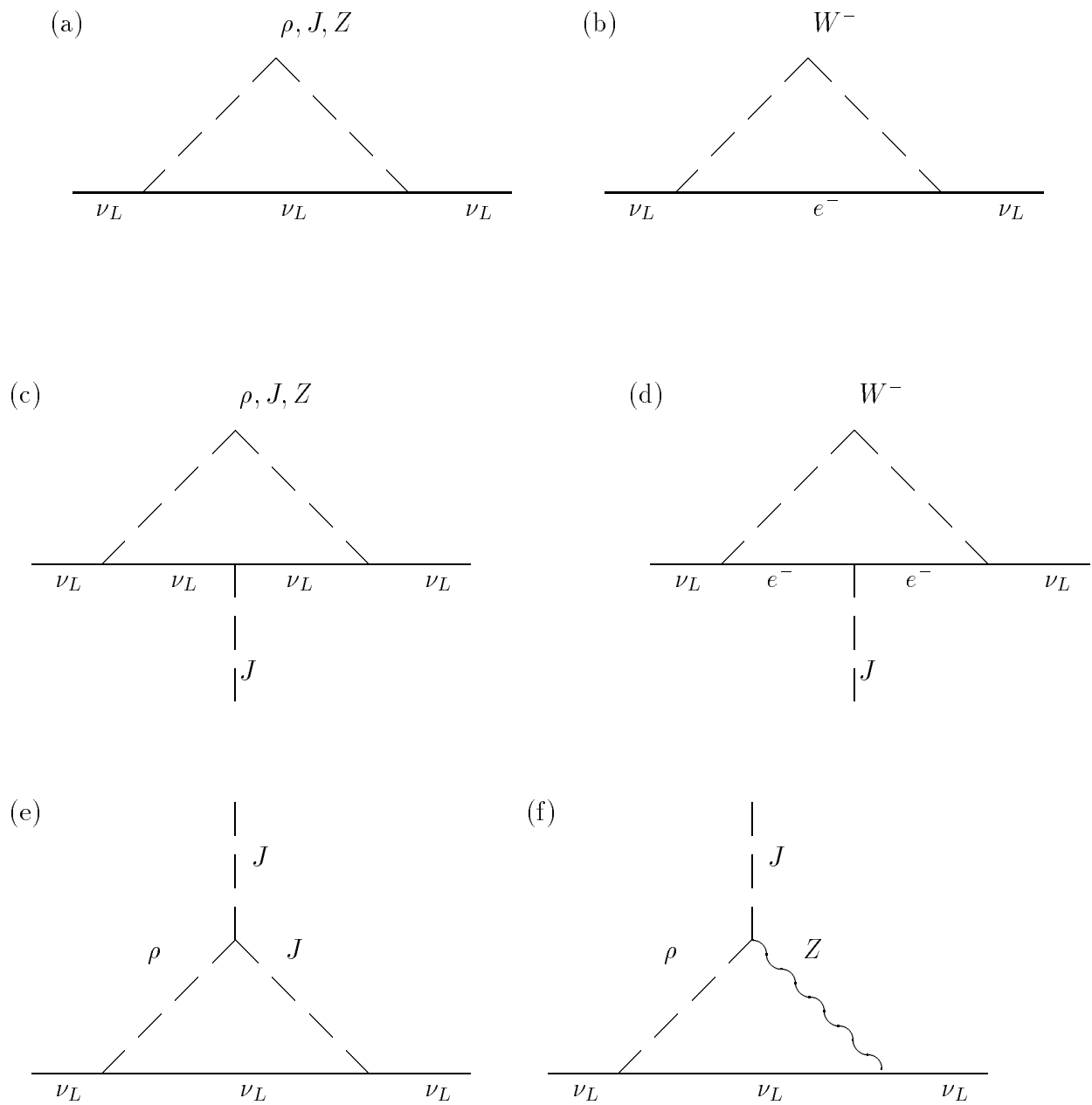
\begin{figure}
\caption{One loop diagrams contributing to the neutrino mass and to the
neutrino-majoron coupling in the
triplet majoron
model.
}\label{kuvaTrip}
\end{figure}

In Ref.~\cite{GeorgiGlashowNussinov81} it was claimed
that the graph \ref{kuvaTrip}a would yield the non-diagonal
couplings. Indeed, it contributes to the majoron couplings,
rotating the neutrino-majoron coupling matrix from its original
basis. However, there are other graphs to be taken into account,
for instance the graph \ref{kuvaTrip}b that contributes to the
mass matrix. One must be very careful with the graphs in
this case, due to the peculiarities of this case. For example, the graph (a)
gives in fact a higher dimension term
$\nu \nu \Delta \Phi \Phi$, so that the majoron coupling
emerges only after $\Phi$ gets a vacuum expectation value,
while the respective mass correction (b) is there even if
$\Phi$ would have zero vacuum expectation value.
Furthermore, the mass correction is divergent so that an
appropriate renormalisation procedure should be applied for
the results to make sense. The renormalisation then can
and should be done in the symmetric phase, so that the
results remain to be valid also after the breaking of the symmetries,
for the shifted fields.

Being now convinced that the argument of current conservation is
sufficient, the calculation of the loops is skipped.
Instead the results obtained in the section
\ref{Noether} by N\"{o}ther's theorem
are applied.
{}From the formula (\ref{nemaj}) one obtains directly
\begin{equation}
\xi = m \frac{v}{u\sqrt{v^2+4u^2}},
\end{equation}
valid after infinite radiative corrections.
Hence the majoron coupling remains strictly diagonal,
contradictory to the claims of ref.~\cite{GeorgiGlashowNussinov81}.

\section{Other models}

\subsection{Singlet-triplet model}

The simplest cure to the triplet majoron model is introducing
a singlet higgs. In the singlet-triplet majoron model the majoron
is given by the composition \cite{SchechterValle82}
\begin{equation}
J = \sqrt{2} \; {\rm Im} \left[
\frac{-2 vu^2 \Phi^o + w(v^2+4u^2)\sigma + uv^2 \Delta^o}
{\sqrt{4v^2u^2 + w^2(v^2+4u^2)+u^2v^4}} \right].
\end{equation}
One can choose the parameters so that the majoron
is predominantly a singlet, thus avoiding the LEP constraint.
The new singlet couples only to the other scalars,
via the coupling
\begin{equation}
\lambda' \sigma \Phi^T \Delta \Phi.
\end{equation}
There will be no new graphs contributing to the mass
or to the majoron coupling of neutrinos, hence the above results
apply, corrected only by some mixing factors. Thus the neutrinos
can be considered absolutely stable in the  singlet-triplet
majoron model with the minimum fermion contents.

Allowing the existence of the right-handed neutrinos would lead
to a case where the neutrino mass is generated mixedly
by the triplet field and by the singlet field through the
see-saw mechanism. It is reminded that despite the two different mechanisms
for the light neutrino mass, the neutrino-majoron coupling remains diagonal at
$\epsilon^2$ \cite{SchechterValle82}. At one loop level the result is
corrected by all the graphs
considered for both the singlet and the triplet majoron model, with
only the scalar mixing factors being more complicated.
There appear also some new graphs due to the mixing of
the majoron with the doublet higgs.
Again, instead of calculating the graphs, the result (\ref{nemaj}) is applied,
giving
\begin{equation}
\xi = \frac{v^2 m + (v^2 + 2 u^2)
(m \epsilon\epsilon^\dagger + \epsilon \epsilon^\dagger m)}
{\sqrt{4v^2u^2 + w^2(v^2+4u^2)+u^2v^4}} ,
\end{equation}
where the non-diagonal parts emerge from the latter terms.

\subsection{Doublet majoron models}

The doublet majoron models \cite{BertoliniSantamaria88}
have also suffered from
the results of LEP.
In the simplest version of it, one adds to the standard model
a new scalar doublet with lepton number 1. Then we need also
three right-handed neutrinos which
carry no lepton number, so that their Majorana mass term can
appear in the bare Lagrangian.
As a result of the spontaneous breaking of the
lepton number (and $SU(2)\times U(1)$ as well)
the neutrinos get Dirac mass terms.
The mass matrix is diagonalised
exactly as in the singlet majoron case, and the light
neutrinos get their masses due to the see-saw mechanism.

Apart from the singlet majoron model, the majoron couples only
between the left- and right-handed neutrinos. Due to the
explicit non-coincidence of the mass matrix and majoron couplings,
the majoron coupling can be non-diagonal even at the tree level.
In the mass basis the majoron coupling is given by
\begin{eqnarray}
\xi &=& \frac{1}{w}\left[
\begin{array}{cc}
\widehat{\xi}& -\epsilon M_M  \\
- M_M \epsilon^\dagger & \breve{\xi}
\end{array}
\right]
+
\left[\begin{array}{cc}
O(\epsilon^6) & O(\epsilon^3) \\
O(\epsilon^3) & O(\epsilon^4)
\end{array}\right]
\end{eqnarray}
where
\begin{eqnarray}
\widehat{\xi} &=&-2\epsilon M_M\epsilon^\dagger +
2\epsilon M_M \epsilon^\dagger\epsilon\epsilon^\dagger
+ 2\epsilon\epsilon^\dagger\epsilon M_M \epsilon^\dagger
\\
\breve{\xi} &=& 2M_M - 2 M_M
\epsilon^\dagger \epsilon
-2 \epsilon^\dagger \epsilon M_M
\end{eqnarray}
Coincidentally, the non-diagonal majoron coupling of the light
neutrinos is given by the same equation as that of the singlet
majoron model (with an insignificant sign difference).
However, in the doublet majoron model (despite the apparently
similar formula) the neutrino majoron coupling
constant is generally larger, since the scale $w$ is supposed
to be lower than in the singlet majoron model.

It is easy to verify explicitly that the result is valid also with
the radiative corrections. All the relevant loops
are well-behaving graphs, structurally similar to those
considered in singlet majoron case, giving together convergent
results. Omitting the explicit proof, one concludes that the loops do
not add any relevant contribution, as follows from
the current conservation.

Recently a variant of the doublet majoron model \cite{KikuchiMa94} was
presented that adds another (third) doublet to the model. This allows the
$\rho$ field to be heavy enough to disallow the decay $Z \to J\rho$,
without invoking a too big majoron coupling to the charged leptons.
Another remedy is to add a singlet scalar. In neither case there
will arise any new contributions up to one loop level, since
the introduced particles do not couple to leptons.

\subsection{Doublet-triplet models}

For completeness,
consider finally the theoretical doublet-triplet model.
In this scenario the majoron consists of a triplet and
a doublet that does not have to be the standard model
doublet.
Since the new doublet does not couple to leptons,
the results of the triplet majoron models apply,
with some trivial corrections due to mixing.
Hence, the majoron is now given by
\begin{equation}
J = \sqrt{2}\; {\rm Im} \left[\frac{ (w^2+4u^2) \Phi^\circ
 - vw \varphi^\circ
- 2 uv \Delta^0}
{\sqrt{(v^2+w^2+4u^2)(w^2+4u^2)}} \right],
\end{equation}
where $\Phi$ denotes the standard scalar doublet,
$\varphi$ the new doublet and $\Delta$ the scalar
triplet, and $v, w$ and $u$ are their vacuum expectation
values, respectively. Its coupling to neutrinos reads
\begin{equation}
\xi = m \frac{2v}{\sqrt{(w^2+4u^2)(v^2+w^2+4u^2)}},
\end{equation}
being truly diagonal.

With this particle contents one can avoid the
inconsistency between $Z$ decay width and $\rho$
parameter, but the model yet fails to satisfy the
astrophysical constraints for any natural choice
of parameters. To cure that, one should add more
higgses, and as long as they do not couple to
leptons, there are no new one-loop graphs to
be taken into account.
In any case the general result guarantees that there
are no new effects due to the higher order corrections.

\section{Majoron coupling to charged fermions}
\label{majel}

It has been often speculated (see \cite{MohapatraPal88,PeltoniemiValle93a}
for a few examples) that a substantial coupling between
majorons and charged leptons would arise by radiative
corrections. Since the phenomenology of the charged leptons
is relatively well understood, the laboratory experiments, as well
as astrophysical considerations, constrain such couplings severely.

We can derive the general majoron coupling to Dirac fermions
using the same method of current conservation as for neutrinos.
Hence we obtain
\begin{eqnarray}\label{cfmaj}
\xi_{f} =
-C
 \left(\sum_h L_h Y_h v_h^2\right) m_f,
\end{eqnarray}
for a fermion f
with the standard quantum number assignments, and a Dirac mass $m_f$.

The above result (\ref{cfmaj}) would imply that in the singlet majoron models
the majoron does not couple to any other known fermion
than neutrinos. Such a conclusion might also appear obvious
from the initial form of the effective coupling,
$\sigma^\dagger \sigma \bar{f}f$.
In models where the majoron is composed
of non-singlet components it couples to fermions already at
the tree level, via its doublet component.
However, according to (\ref{cfmaj}), the majoron coupling should be always
diagonal in all models with the standard
fermion sector, and hence
the decay $f \to f' J$  should be forbidden.

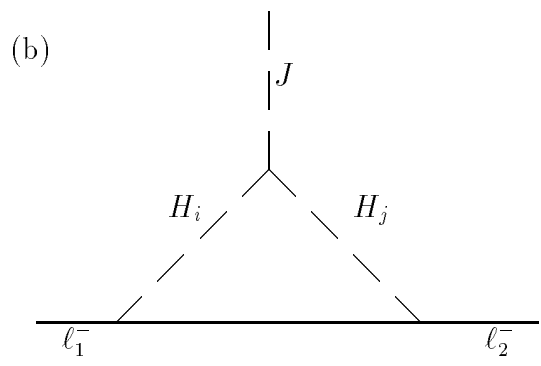
\begin{figure}
\caption{Diagrams apparently contributing to the coupling of
the majoron to the charged leptons, in a model where a singlet majoron couples
directly only to scalars. These graphs cancel.
}
\label{kuvaJc}
\end{figure}

Let us now consider as an example a specific model in more detail.
The simplest case is a model with a singlet majoron that does not
couple directly to fermions, but has a coupling with other scalars,
introduced e.g. to generate radiative mass terms. Hence we assume
three new scalars, $\chi_1$, $\chi_2$ and $\sigma$, where
$\chi$ can be doubly charged scalars, as in \cite{PeltoniemiValle93a},
or some other representation. Now the loops depicted in the figure
\ref{kuvaJc} would apparently induce a coupling of the majoron
to the charged leptons. However, the coupling $J \chi_i \chi_j$
is antisymmetric, from which it directly follows that the graphs
with different scalar mass eigenstates cancel. Hence the majoron
coupling to the charged lepton vanishes in this model.

\section{Conclusions}

It was claimed that in models with
spontaneously broken lepton number symmetries
the radiative
 corrections do not enhance the invisible decay rates of neutrinos. The
graphs with external majorons are always
accompanied with similar graphs without
majorons that contribute to the mass, and the resulting mass and majoron
coupling matrices  turn out to be simultaneously
diagonal.

It was explicitly demonstrated that in the singlet majoron model
 the dominant component to neutrino
decay appears at tree level, and the radiative corrections do not
enhance it.

It was found that the coincidence between the neutrino mass and
the neutrino-majoron coupling is maintained also in models
where both are generated radiatively. The neutrino-majoron
coupling at one loop level was calculated exactly in the Zee
model with a singlet majoron.  It is evident that the result applies to other
radiative models with similar ingredients.

A general proof for the coincidence between mass matrix and neutrino
majoron coupling was given using N\"o{}ther's theorem.
This result is valid for all well-behaving theories up to infinite
radiative corrections.

We can conclude that the neutrino-majoron coupling is strictly
diagonal, and the neutrinos are stable against a decay
to a majoron in models where all neutrinos have identical lepton
number and hypercharge. This is the case in all models with the
minimal fermion content and a canonical lepton number assignment.
Examples include triplet and doublet majoron models and their derivatives not
containing the   right-handed neutrinos.

The general arguments apply also to the coupling of majorons to
other fermions. Hence in models where the majoron has a doublet component
it couples to charged fermions at tree level, and the radiative corrections
do not induce any non-diagonal components.
In models where the majoron is made of singlets it does not couple at all
to charged leptons or quarks having conventional quantum numbers.
It was shown
explicitly that in a specific singlet majoron model the
one-loops apparently
contributing to such a coupling vanish. These results do not necessarily
apply for models with
non-canonical lepton number assignments and exotic fermions.

The above arguments do not forbid higher dimension terms involving many
majorons, like $\bar{f}f JJ$, that could be induced radiatively.
Nevertheless even though such terms would appear, they would be very small,
especially the non-diagonal components. Moreover, the decay
$f_2 \to f_1 J^n$ would be suppressed by phase space arguments.

It has been speculated that the global symmetry may be broken
both explicitly and spontaneously \cite{GelminiNussinovRoncadelli82}.
This might destabilize neutrinos even in models where they are otherwise
stable, if the majoron remains sufficiently massless. For instance, in the
triplet majoron model, a small gravitationally induced neutrino mass of
O($10^{-5}$) eV is sufficiently big to cause any neutrino considered
relevant for the dark matter to decay in time scales much less than the
age of the universe.  On the other hand, in typical singlet majoron
models the majoron coupling is so small that the gravitationally induced
mass is not sufficient to enhance the decay rate.

The explicit symmetry breaking may also induce a mass to the majoron.
Its mass could be as high as tens of keV in which case one does not talk
about the decay of light neutrinos,
even if the explicit symmetry breaking terms
might induce sizable non-diagonal couplings. Many astrophysical constraints
do not apply for such majorons which may give more freedom to models.

\section*{ ACKNOWLEDGEMENT}

This work was supported by the
Spanish Ministry for Education and Science, and by Istituto Nazionale di
Fisica Nucleare.
I have benefitted from the discussions with Kimmo Kainulainen, Palash Pal and
Jos\'e Valle.

\appendix

\section{The absence of the $\nu_L \nu_R J$-
coupling in the singlet majoron model}

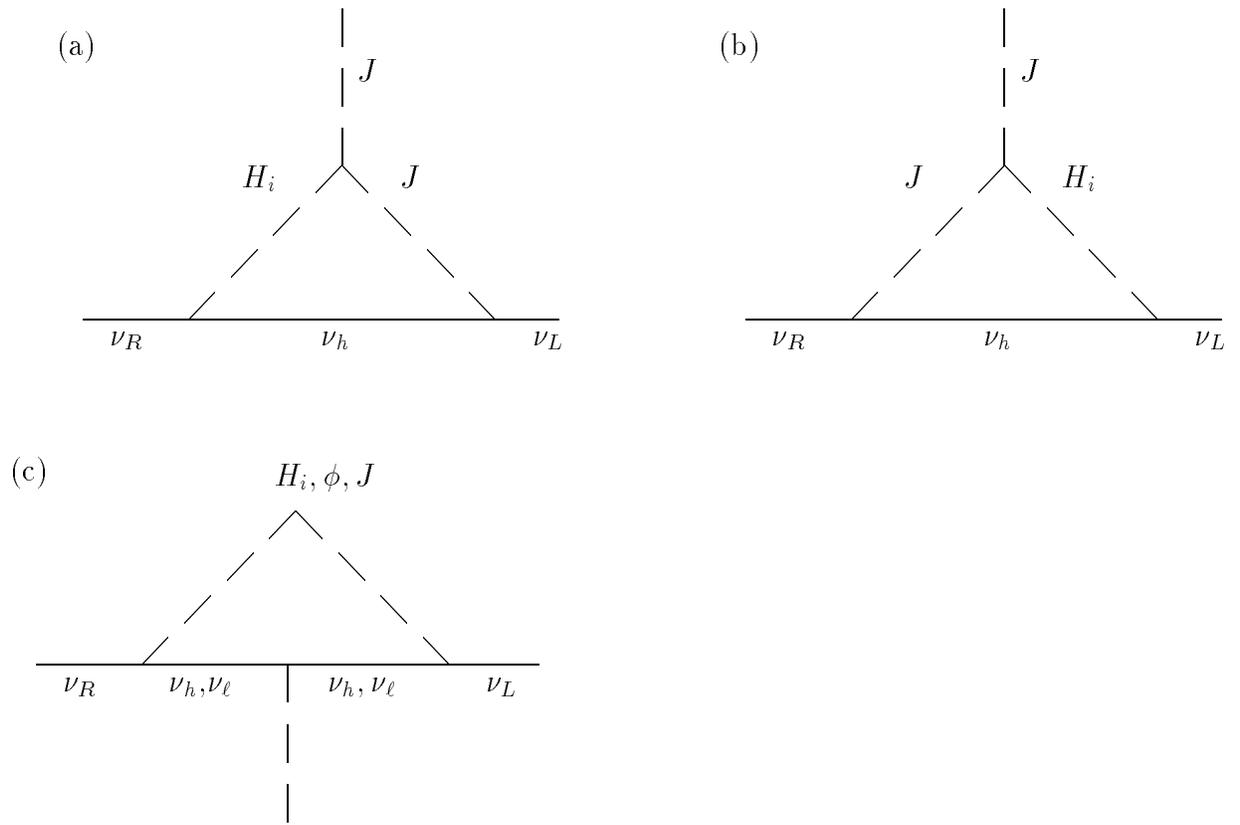
\begin{figure}
\caption{The would-be diagrams contributing to the majoron coupling between
left- and right-handed states. These graphs cancel.
}
\label{kuvaLRJ}
\end{figure}

There was no majoron coupling between the left and right handed components at
the tree level; neither will it arise by
loops which is now shown explicitly, calculating the graphs of figure
\ref{kuvaLRJ}.
Summing all the fermionic contributions
together, we obtain, for the $\phi$ exchange graph
\begin{eqnarray}
-i \gamma_5 \delta \xi_{ai}^{(\phi)} &=&
\sum_{bj} i
\int \frac{f_{aj}\gamma_5}{\sqrt{2}}\epsilon_{jb}
i\frac{\backslash \hspace{-0.5em}{k}+\check{m}_b}{k^2-\check{m}_b^2 }
\frac{-h_b \gamma_5}{\sqrt{2}}
i\frac{\backslash \hspace{-0.5em}{k}-
\backslash \hspace{-0.5em}{q}+\check{m}_b}{(k-q)^2-\check{m}_b^2 }
\nonumber \\ &&
\frac{i}{(p-k)^2-M_Z^2 }\frac{f_{bi}\gamma_5}{\sqrt{2}}
\frac{{\rm d}^4 k}{(2\pi)^4 }
\nonumber \\ && \mbox{}+
\sum_{j \ell k b} i
\int \frac{f_{aj}\gamma_5}{\sqrt{2}}U_{j\ell}
i\frac{\backslash \hspace{-0.5em}{k}+\hat{m}_\ell}
{k^2-\hat{m}_\ell^2 }U^*_{\ell k}
(-\epsilon_{kb})
\frac{-h_b \gamma_5}{\sqrt{2}}
\nonumber \\ &&
i\frac{\backslash \hspace{-0.5em}{k}-
\backslash \hspace{-0.5em}{q}+\check{m}_b}{(k-q)^2-\check{m}_b^2 }
\frac{i}{(p-k)^2-M_Z^2 }\frac{f_{bi}\gamma_5}{\sqrt{2}}
\frac{{\rm d}^4 k}{(2\pi)^4 }.
\end{eqnarray}
One sees immediately that the sum vanishes (trivially when the light
neutrino masses $m_\ell$ are ignored, at order $m_\ell$ more accuracy is
required).
Respectively vanishes the
contribution by the massive higgses.
This is expected, since this coupling does
not break the lepton number symmetry. It can be seen clearly also in the
perturbative
picture: the $\sigma$ lines appear always pairwise,
one coming in and another
going out.

Due to the scalar mixing there will arise also new type of graphs as
those shown in figure \ref{kuvaLRJ}.
The new contribution to the
Dirac mass is
\begin{eqnarray}
\delta  m_{ai} =
\frac{h_a \check{m}_a f_{ai}}{32 \pi^2 } \sin 2\theta
(M_1^2-M_2^2) I(\check{m}_a,M_1,M_2).
\end{eqnarray}
There are two sets of graphs contributing to the majoron coupling.
Those with the majoron in the fermion leg  read
\begin{eqnarray}
\delta  \xi_{ai}^{(a)} =
\frac{h_a h_a f_{ai}}{64 \sqrt{2} \pi^2 } \sin 2\theta
(M_2^2-M_1^2) I(\check{m}_a,M_1,M_2).
\end{eqnarray}
Another set consists of graphs where the majoron touches the scalar loop.
The effective cubic couplings of a massive scalar to two majorons
($D_i H_i JJ $),
can be written explicitly in the mass basis of the neutral scalars as
\begin{eqnarray}
D_{J1} &=& -\frac{h_a}{\sqrt{2} \check{m}_a} M_1^2 \sin \theta
\nonumber \\
D_{J2} &=& \frac{h_a}{\sqrt{2} \check{m}_a} M_2^2 \cos \theta,
\end{eqnarray}
where $a$ is arbitrary.
Hence one obtains
\begin{eqnarray}
\xi_{ai}&=&
\frac{h_a f_{ai} h_a}{2\sqrt{2}}
\sin \theta \cos\theta
\frac{1}{16\pi^2}
\nonumber \\&&
[M_1^2 I(\check{m}_a,0,M_1) -M_2^2 I(\check{m}_a,0,M_2)]
\end{eqnarray}
Now, using the identity
\begin{eqnarray}
 \lefteqn{(M_2^2-M_1^2) I(\check{m}_a,M_1,M_2) = } & &  \nonumber \\
& & M_2^2 I(\check{m}_a, 0, M_2) - M_1^2 I(\check{m}_a, 0, M_1)
\end{eqnarray}
one sees that these loops cancel, so that the result is zero.
Hence we find that the mixing of the scalars does not induce any
left-right components for the majoron coupling matrix.



\setlength{\itemsep}{0mm}

\clearpage

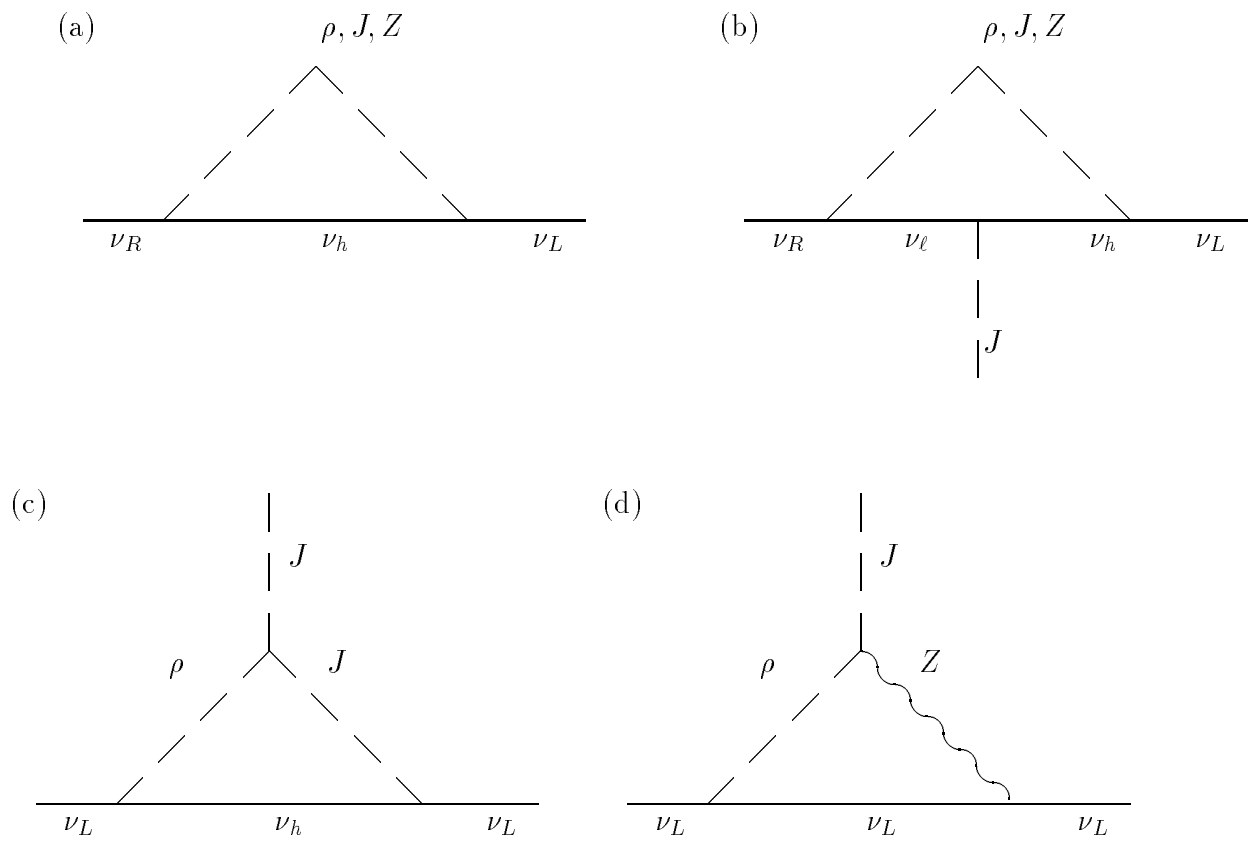
\begin{figure}
\caption{The most important one-loop diagrams contributing to the neutrino mass
and the
neutrino-majoron coupling in the
doublet majoron
model.
}\label{kuvaDoub}
\end{figure}

\end{document}